\documentclass{PoS}

\def\3{\ss{}}

\def\eps{\epsilon}

\def\lg{\left\{ }
\def\rg{\right\} }

\newcommand{\beq}{\begin{equation}}
\newcommand{\eeq}{\end{equation}}
\newcommand{\bea}{\begin{eqnarray}}
\newcommand{\eea}{\end{eqnarray}}
\newcommand{\bfig}{\begin{figure}}
\newcommand{\efig}{\end{figure}}
\newcommand{\bc}{\begin{center}}
\newcommand{\ec}{\end{center}}
\newcommand{\dd}{\mathrm{d}}

\def\to{\rightarrow}

\def\eps{\epsilon}
\def\mgl{m_{\tilde g}}
\def\msq{m_{\tilde q}}
\def\msqc{m_{\tilde{q}_c}}
\def\mga{m_{\tilde \chi}}

\title{Precision predictions for associated gluino-gaugino production at the LHC}

\ShortTitle{Precision predictions for associated gluino-gaugino production at the LHC}

\author{Benjamin Fuks%
 \thanks{Supported by CNRS under contract PICS 150423 and Th\'eorie-LHC-France initiative of CNRS (IN2P3/INP).}\\
 Sorbonne Universit\'es, UPMC Univ.~Paris 06, UMR 7589, LPTHE, F-75005 Paris, France\\
 CNRS, UMR 7589, LPTHE, F-75005 Paris, France\\
 Institut Universitaire de France, 103 boulevard Saint-Michel, F-75005 Paris, France \\
 E-mail: \email{fuks@lpthe.jussieu.fr}}

\author{\speaker{Michael Klasen}
 \thanks{Supported by BMBF under contract 05H15PMCCA.}\\
 Institut f\"ur Theoretische Physik, Westf\"alische
 Wilhelms-Universit\"at M\"unster, Wilhelm-Klemm-Stra\ss{}e 9,
 D-48149 M\"unster, Germany\\
 E-mail: \email{michael.klasen@uni-muenster.de}}

\author{Marthijn Sunder%
 \thanks{Supported by DFG under contract GRK 2149.}\\
 Institut f\"ur Theoretische Physik, Westf\"alische
 Wilhelms-Universit\"at M\"unster, Wilhelm-Klemm-Stra\ss{}e 9,
 D-48149 M\"unster, Germany\\
 E-mail: \email{mpasunder@uni-muenster.de}}

\abstract{
\vspace*{-160mm}
\flushright{\large MS-TP-17-12}\\
\vspace*{150mm}
 Now that the mass limits for gluinos have been pushed to the few-TeV
 range, they might only be visible at the LHC in associated production with
 lighter gauginos. We compute the corresponding cross section at
 next-to-leading logarithmic (NLL) and next-to-leading order (NLO) precision
 in the QCD coupling constant. The resulting expressions are implemented
 in the public code RESUMMINO and can be directly used in the corresponding
 experimental searches.}

\FullConference{EPS-HEP 2017, European Physical Society conference on High Energy Physics\\
		5-12 July 2017\\
		Venice, Italy}

\begin{document}

\section{Cross sections for supersymmetric particles at the LHC}

Supersymmetry (SUSY), a long-standing, well-motivated and complete extension
of the Standard Model (SM) of particle physics with a rich phenomenology,
continues to be searched
for at CERN's Large Hadron Collider (LHC). For many years, squarks and gluinos
have been at the centre of this search due to their strong interactions and
correspondingly large cross sections. However, the mass limits for these
particles are now already in the few-TeV range, so that pair production of
these particles might soon get out of reach at LHC energies of 13 or 14 TeV.
Associated production of squarks and gluinos with an electroweak superpartner
(a gaugino or higgsino, or equivalently a neutralino or chargino), that is
still allowed to be light and that is motivated for its part by dark matter
observations, might then be the only possibility to study them. The
corresponding cross sections, which
are, like the average produced final-state mass, of intermediate size and have
been known for many years at next-to-leading order (NLO) of QCD, should then
be known with the same up-to-date precision as those for strong and weak SUSY
particle pair production, i.e.\ at next-to-leading logarithmic (NLL) accuracy
\cite{Fuks:2013vua} and beyond \cite{Beenakker:2016lwe}, so that they can
be used by the ATLAS \cite{Aad:2015eda} and CMS \cite{Khachatryan:2014qwa}
collaborations.

\section{Analytical results}

The tree-level Feynman diagrams for associated gluino-gaugino production
\begin{figure}[ht]
\begin{center}
 \includegraphics[width=0.4\textwidth]{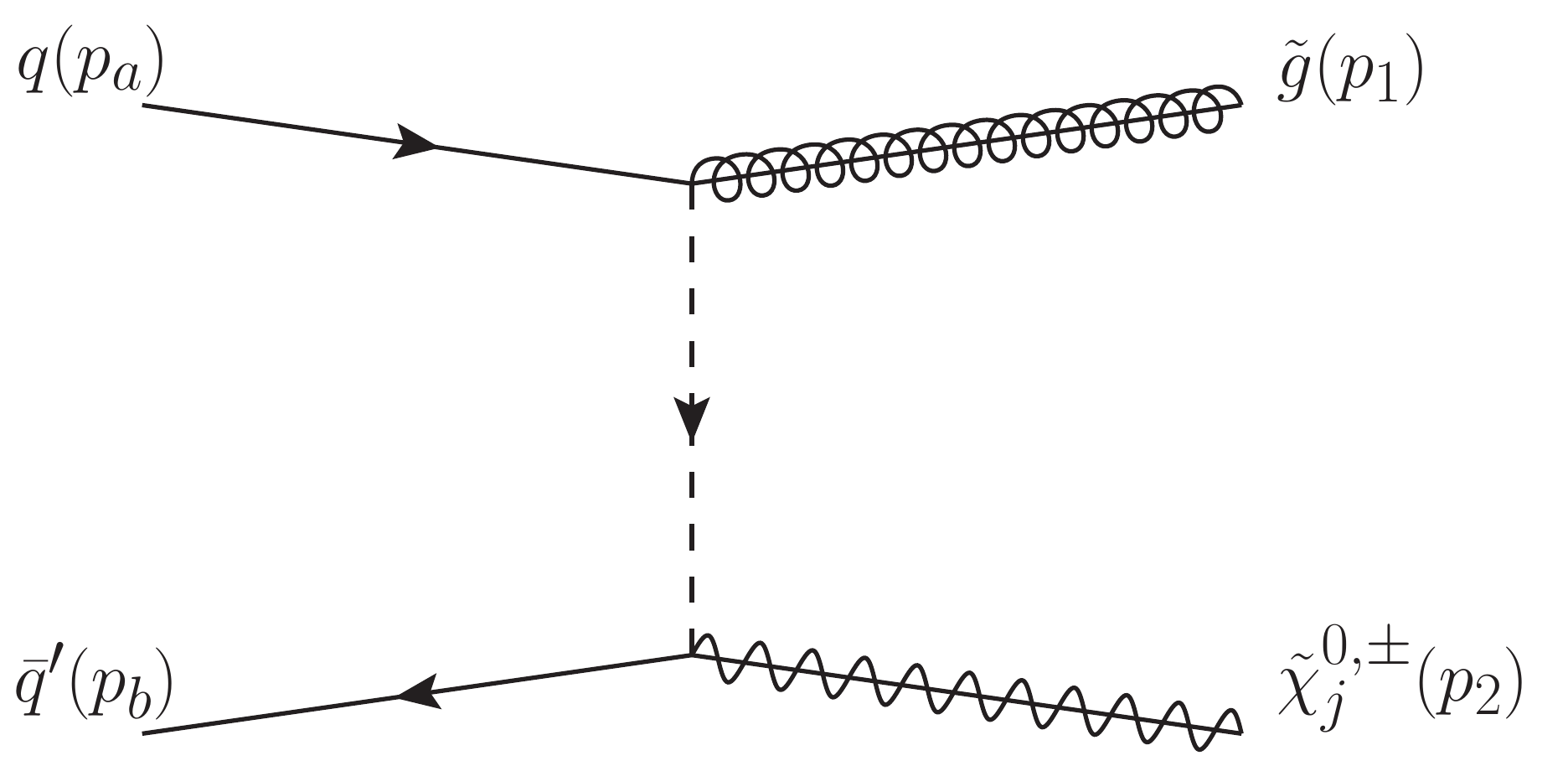}
 \includegraphics[width=0.4\textwidth]{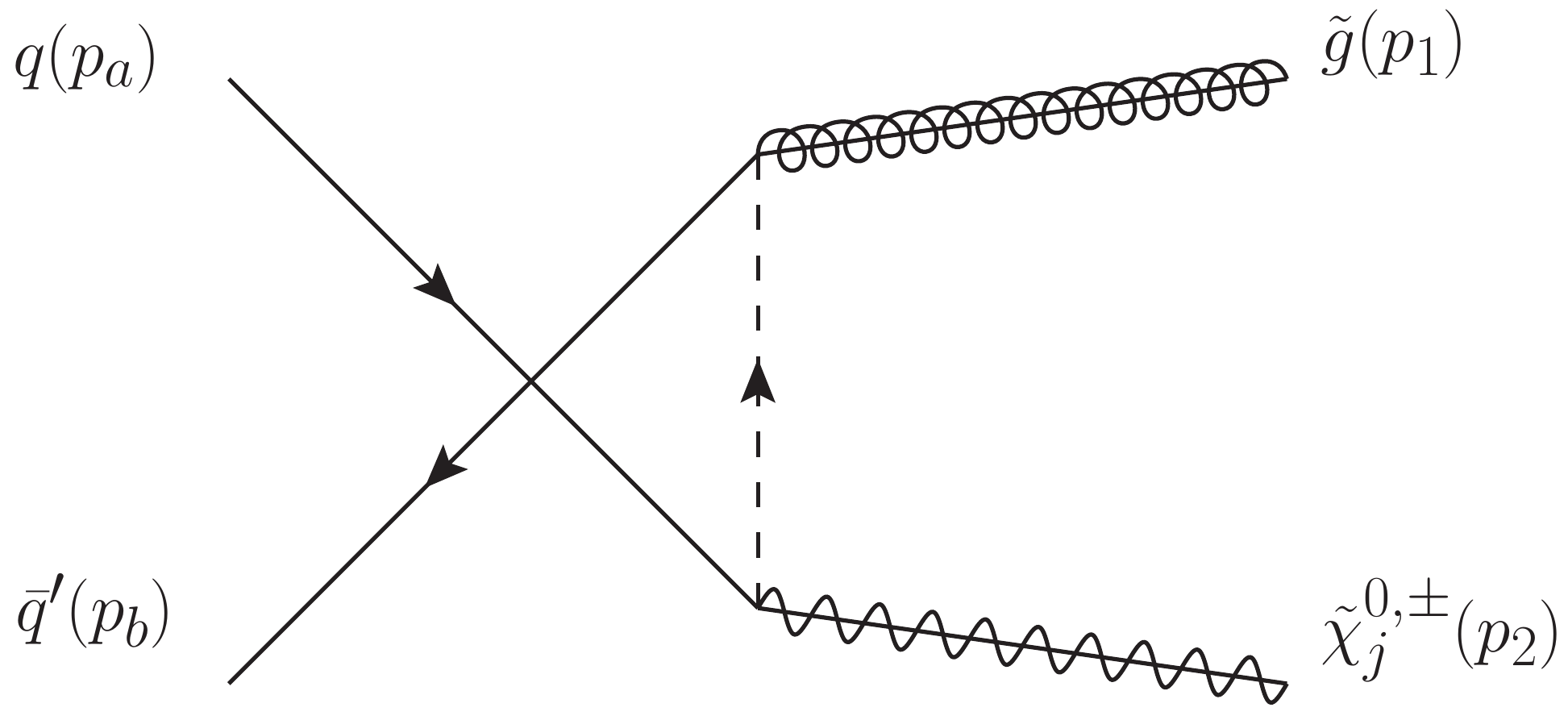}
 \caption{Tree-level Feynman diagrams for associated gluino-gaugino production.}
 \label{fig:01}
\end{center}
\end{figure}
are shown in Fig.~\ref{fig:01}.
They lead to the squared matrix elements
\bea
 \mathcal{M}_t\mathcal{M}^*_{t_c} &=& \frac{C_A C_F\,e\,g_s(\mu_r)}{(\msq^2 - t) (\msqc^2 - t)} ( \mathcal{L}' \mathcal{L}'_c + \mathcal{R}' \mathcal{R}'_c) (L L_c + R R_c)(\mgl^2 - t) (\mga^2 - t),
\\
 \mathcal{M}_u\mathcal{M}^*_{u_c} &=& \frac{C_A C_F\,e\,g_s(\mu_r)}{(\msq^2 - u) (\msqc^2 - u)}( \mathcal{L} \mathcal{L}_c + \mathcal{R} \mathcal{R}_c) (L' L'_c + R' R'_c)(\mgl^2 - u) (\mga^2 - u),
\\
\mathcal{M}_t \mathcal{M}^*_{u_c} &=& \frac{C_A C_F\,e\,g_s(\mu_r)}{(\msq^2 - t)(\msqc^2 - u)} \biggl[ \left(-s^2 + t^2 + u^2 + (\mga^2 + \mgl^2 ) (s - t - u) + 2 \mgl^2 \mga^2\right) \nonumber \\
&\times& \left( L L_c \mathcal{L}' \mathcal{L}'_c + R R_c \mathcal{R}' \mathcal{R}'_c \right) 
+ 2 \mgl \mga s (\mathcal{R} R_c  \mathcal{L}' L'_c   + \mathcal{L} L_c   \mathcal{R}' R'_c) \biggr],
\eea
and consequently the partonic and hadronic cross sections
\beq
 \dd\sigma_{ab}^{(0)}=\int_2 \dd \sigma^{B} = \int \frac{1}{2s}\,
 \frac{1}{4C_A^2} \sum_{{\tilde{q},\tilde{q}_c}} \left(\mathcal{M}_t\mathcal{M}^*_{t_c} + \mathcal{M}_u\mathcal{M}^*_{u_c} - 2 {\rm Re}(\mathcal{M}_t \mathcal{M}_{u_c}^*) \right)
% \overline{|{\cal M|}^2}
\, {\rm dPS}^{(2)}
\eeq
and
\bea
 \sigma_{AB}~=~\int M^2\frac{\dd\sigma_{AB}}{\dd M^2}(\tau)&=&\sum_{a,b} \int_0^1 \!
 \dd x_a \,\dd x_b \,\dd z[x_a f_{a/A}(x_a,\mu_f^2)] [x_b f_{b/B}(x_b,\mu_f^2)] \nonumber\\
 &\times& \,[z\,\dd\sigma_{ab}(z,M^2,\mu_r^2,\mu_f^2)]\,\delta(\tau-x_ax_bz).
\eea
Here $C_{A,F}$ denote QCD colour factors, $e,g_s(\mu_r)$ electromagnetic
and (scale-dependent) strong couplings, $s,t,u$ Mandelstam variables,
dPS$^{(2)}$ the two-particle phase space, $L,R$ and $\mathcal{L,R}$ gaugino
and gluino coupling strengths, and $\tau=M^2/S$ the ratio of the squared
invariant mass of the produced SUSY particle pair to the hadronic
centre-of-mass energy.
The NLO corrections are computed with the Catani-Seymour dipole subtraction
method \cite{Catani:2002hc}
\beq
 \dd \sigma^{(1)}_{ab}=\sigma^{\{3\}}+\sigma^{\{2\}}+\sigma^C=
 \int_3[\dd\sigma^{R}-\dd\sigma^A]_{\eps=0}+\int_2[\dd\sigma^V+\int_1\dd\sigma^A]_{\eps=0}
 +\sigma^C
\eeq
in $D=4-2\eps$ dimensions, and the final result agrees with our
previous calculation \cite{Berger:2000iu}.

Close to partonic threshold, when $z={M^2\over s} \to 1$, large logarithms
$\left({\alpha_s\over2\pi}\right)^n \left[{\ln^m(1-z)\over1-z}\right]_+$
spoil the convergence of the perturbative series and must be resummed to all
orders. This is most easily achieved in Mellin space, where the resummed cross
section
\beq
 \dd\sigma^{\rm (res.)}_{ab\to ij}(N,M^2,\mu^2) = \sum_{I}\,\mathcal{H}_{ab\to ij,I}(M^2,\mu^2)\,
 \Delta_a (N,M^2,\mu^2)\,\Delta_b (N,M^2,\mu^2) \Delta_{ab\to ij,I}(N,M^2,\mu^2)
\eeq
factorises into soft-collinear and soft functions
\beq
 \Delta_a\Delta_b\Delta_{ab\to ij,I} = \exp\Big[L G^{(1)}_{ab}(\lambda)
 + G^{(2)}_{ab\to ij,I}(\lambda,M^2/\mu^2) %+ a_s G^{(3)}_{ab\to ij,I}(\lambda,M^2/\mu^2)
 + \ldots \Big]
\eeq
with leading and next-to-leading logarithms \cite{Catani:2003zt}
\bea
G^{(1)}_{ab}(\lambda) &=& g_a^{(1)}(\lambda) + g_b^{(1)}(\lambda), \\
G^{(2)}_{ab \rightarrow ij}(\lambda) &=& g_a^{(2)}(\lambda,M^2,\mu_r^2, \mu_f^2) + g_b^{(2)}(\lambda,M^2,\mu_r^2, \mu_f^2) + h^{(2)}_{ab \rightarrow ij,I}(\lambda).
%G^{(3)}_{ab \rightarrow ij}(\lambda) &=& g_a^{(3)}(\lambda,M^2,\mu_r^2, \mu_f^2) + g_b^{(3)}(\lambda,M^2,\mu_r^2, \mu_f^2) + h^{(3)}_{ab \rightarrow ij,I}(\lambda, M^2, \mu_r^2)
\eea
The soft anomalous dimension $h^{(2)}_{ab \rightarrow ij,I}$ and hard matching
coefficient $\mathcal{H}_{ab\to ij,I}$ are process-dependent. The former reads
in the present case
\beq
 h^{(2)}_{ab \rightarrow ij,I}(\lambda) =
 \frac{2\pi}{\alpha_s}
 \frac{\ln{\left(1 - 2 \lambda\right)}}{2\beta_0}
% D^{(1)}_{ab \rightarrow ij,I}\,,
% D^{(1)}_{ab \rightarrow ij,I} &=&
 {\rm Re} \lg %(\bar{\Gamma}_{ab \rightarrow ij,II})
% {\bar \Gamma}_{ab \rightarrow ij,IJ} = \Gamma_{ab \rightarrow ij,IJ} - \frac{\alpha_s}{2 \pi}
% \sum \limits_{k = {\{a,b\}}} C_k \left( 1-\ln{\left(2 \frac{(v_k \cdot n)^2}{\vert n \vert^2}\right)} - i \pi \right) \delta_{IJ},
% {\bar\Gamma}_{q\bar{q} \rightarrow \tilde{g} \tilde{\chi}} =
 \frac{\alpha_s}{2\pi} C_A \left[\ln{2} + i \pi - 1
 + \ln{\left( \frac{\mgl^2 - t} {\sqrt{2}\mgl\sqrt{s}}\right)}
 + \ln{\left( \frac{\mgl^2 - u }{\sqrt{2} \mgl \sqrt{s}}\right)}\right]
 \rg,
\eeq
while the latter ($\mathcal{H}_{ab\to ij,I}$) can be found in Ref.\ \cite{Fuks:2016vdc}.

\section{Numerical results}

\begin{figure}[ht]
\begin{center}
 \includegraphics[width=0.66\textwidth]{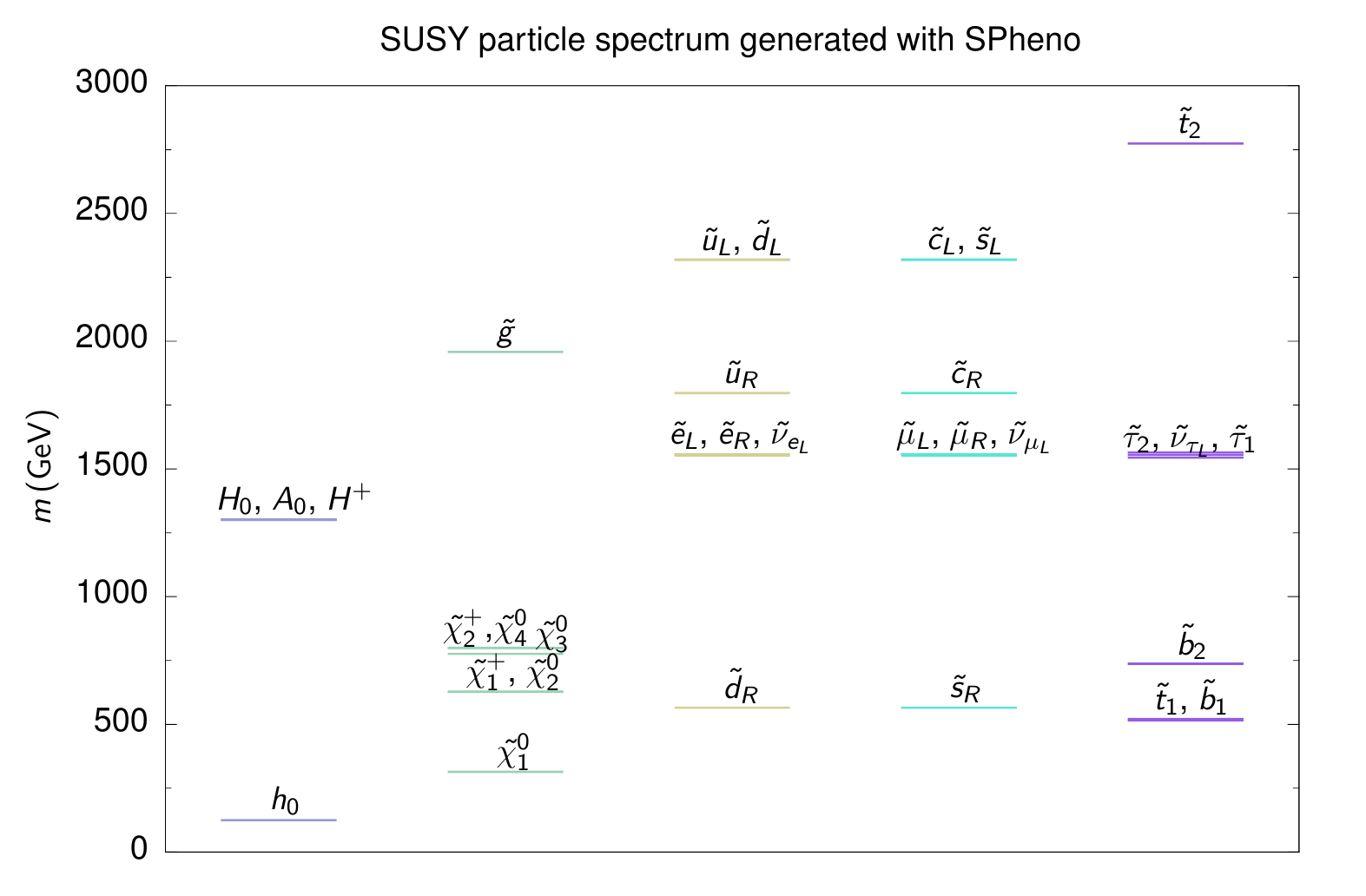}
 \caption{Default pMSSM-13 scenario with light gauginos, a heavy gluino
 and the correct mass of the lightest Higgs boson.}
 \label{fig:03}
\end{center}
\end{figure}

\begin{figure}[ht]
\begin{center}
 \includegraphics[width=0.66\textwidth]{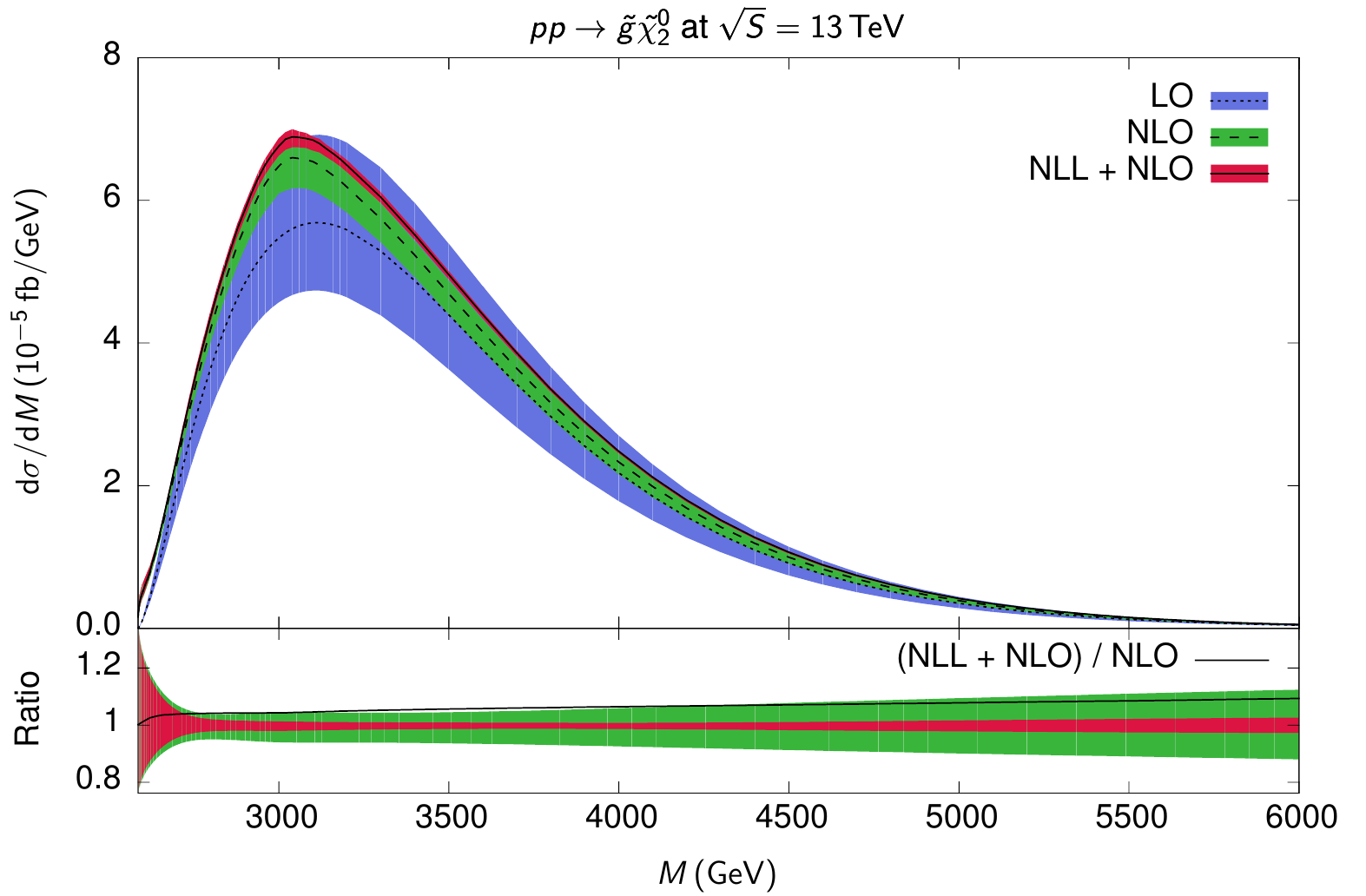}
 \caption{Invariant-mass distributions in LO, NLO and NLL+NLO at the LHC
 with its current centre-of-mass energy of 13 TeV.}
 \label{fig:04}
\end{center}
\end{figure}

\begin{figure}[ht]
\begin{center}
 \includegraphics[width=0.66\textwidth]{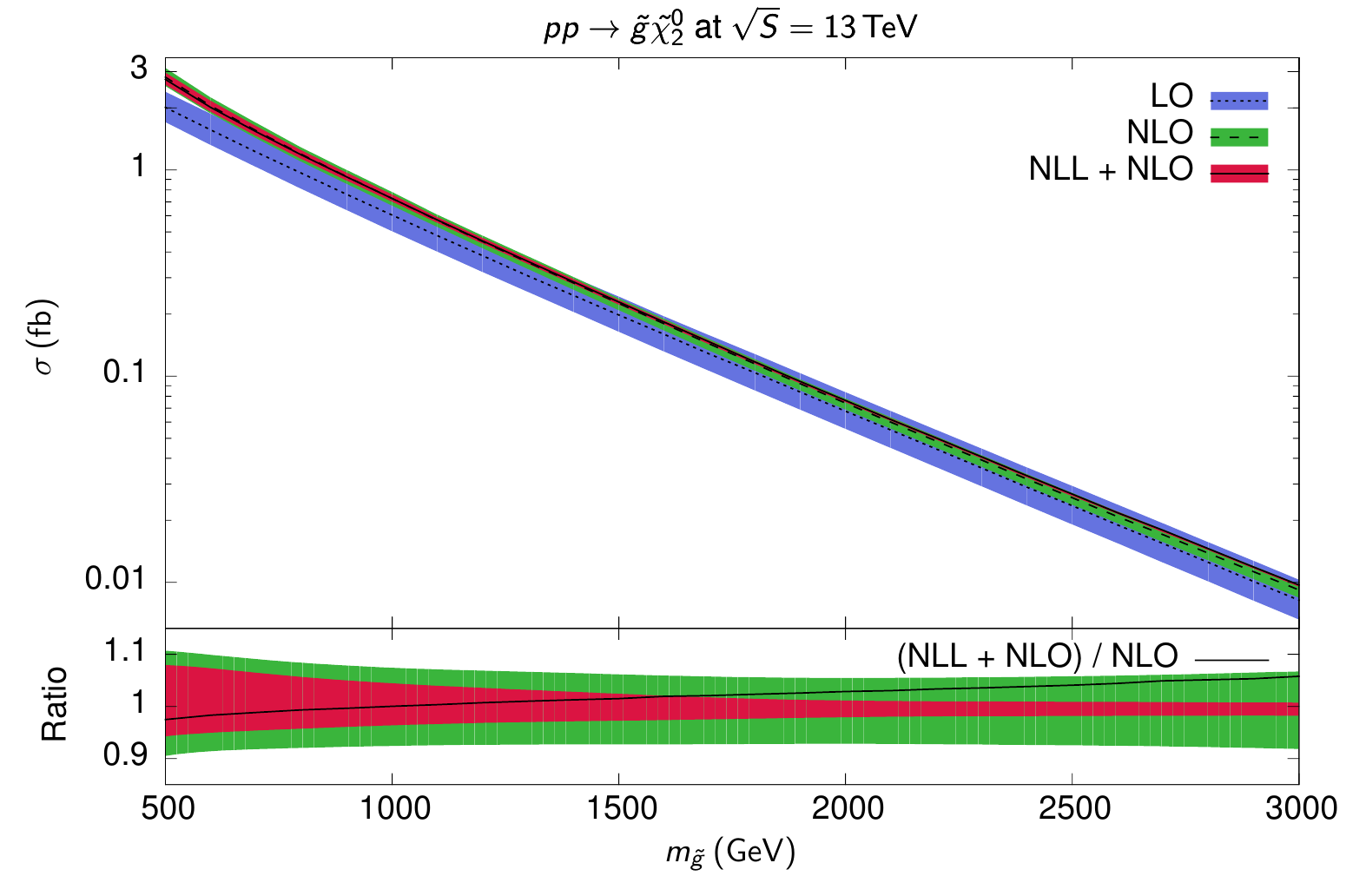}
 \caption{Total cross section as a function of the gluino mass in
 LO, NLO and NLL+NLO.}
 \label{fig:06}
\end{center}
\end{figure}

We study the impact of the NLL+NLO corrections in a phenomenological
Minimal SUSY SM with 13 free parameters (pMSSM-13). Input parameters
are in particular the bino and gluino soft SUSY-breaking masses $M_1$
and $M_3$, with the wino mass fixed to $M_2\simeq 2 M_1$. The
physical mass spectrum is then obtained with SPheno 3.37 \cite{Porod:2011nf}
and shown for our example scenario in Fig.~\ref{fig:03}.
Fig.~\ref{fig:04} shows the invariant-mass distribution of the produced
sparticles. Additional radiation shifts the maximum to lower invariant
masses and reduces the scale dependence. At large $M$, the NLL corrections
increase the NLO cross section by up to 10\%, and the total scale dependence
is reduced there from 30\% to 5\% (lower panel).
The variation of the gluino mass in Fig.~\ref{fig:06} demonstrates that
associated gluino-gaugino pairs in this class of SUSY scenarios will soon be
observable at the LHC with an integrated luminosity of 100 fb$^{-1}$ up to
gluino masses of 3 TeV.

\section{Conclusion}

The semi-weak associated production of gluinos and gauginos might
soon become relevant, if gluinos are too heavy to be pair-produced
at the LHC. This would indeed be theoretically expected from the GUT
relation $M_1=M_2/2=M_3/6$ among the soft SUSY-breaking gaugino and
gluino mass parameters. We have summarised the analytical calculations of
the process-dependent pieces of a threshold-resummation calculation
at NLL accuracy, i.e.\ of the soft anomalous dimension and the hard
matching coefficient, before matching the result to a full NLO calculation
and performing numerically an inverse Mellin transform. As we have
seen, the NLL contributions can increase the NLO invariant mass distribution
by up to 10\% and reduce the total scale dependence from 30\% to 5\%.
The parton density function (PDF) uncertainty should be reduced by
including threshold-improved PDFs in the near future.

\end{document}